%% file: driverFile_arx.tex
\documentclass[12pt]{article}
\pdfoutput=1
\usepackage{epsfig}
\usepackage{subfigure}
\usepackage{rotating}
\usepackage{afterpage}
\usepackage{amsmath} 
\usepackage{amssymb}
\usepackage{relsize}
\usepackage{array}
\usepackage{bm}       
\usepackage{graphicx}

\input workshopsymbols.tex      

\def\babar{\mbox{\sl B\hspace{-0.4em} {\small\sl A}\hspace{-0.37em} \sl B\hspace{-0.4em} {\small\sl A\hspace{-0.02em}R}}}

\begin{document}

\hfill CALT 68-2861
\bigskip


\begin{center}
{\Large\bf The next generation of $\mu \to e \gamma$ and  $\mu \to 3e$\break
CLFV search experiments\break\\}
{\large C.-h. Cheng, B. Echenard and D.G. Hitlin\break\\}
\vskip -12pt
{\large California Institute of Technology, Pasadena, California 91125, USA}
\end{center}
\smallskip
\begin{abstract}
We explore the possibilities for extending the sensitivity of current searches for the charged lepton flavor violating decays $\mu\to e \gamma$ and  $\mu\to eee$.  A future facility such as Project X at 
Fermilab could provide a much more intense stopping $\mu^+$  beam, facilitating more sensitive searches, but improved detectors will be required as well. Current searches are limited by accidental and physics backgrounds, as well as by the total number of stopped muons. One of the limiting factors in current detectors
for $\mu \to e \gamma$ searches is the photon energy resolution of the calorimeter. We present a new fast Monte Carlo simulation of a
conceptual design of a new experimental concept that
detects converted $e^+e^-$ pairs from signal photons, taking advantage of the  improved
energy resolution of a pair spectrometer based on a silicon charged particle tracker.  We also study
a related detector design for a next generation $\mu\to eee$ search experiment.
\end{abstract}

\input introduction_arx.tex

\input mutoegamma_arx.tex

\input mutoeee_arx.tex

\input conclusion.tex

\input bibliography.tex
\end{document}

%% file: workshopsymbols.tex
\newcommand{\nc}{\newcommand}  

\def\beq{\begin{equation}}
\def\eeq#1{\label{#1}\end{equation}}
\def\eeqn{\end{equation}}

\newenvironment{Eqnarray}%
   {\arraycolsep 0.14em\begin{eqnarray}}{\end{eqnarray}}
\def\beqa{\begin{Eqnarray}}
\def\eeqa#1{\label{#1}\end{Eqnarray}}
\def\eeqan{\end{Eqnarray}}

\nc{\ra}{\rightarrow}  
\nc{\slsh}{\slash\hspace*{-0.22cm}}
\def\Re{{\cal R \mskip-4mu \lower.1ex \hbox{\it e}\,}}
\def\Im{{\cal I \mskip-5mu \lower.1ex \hbox{\it m}\,}}

\nc{\vev}[1]{ \left\langle {#1} \right\rangle }
\nc{\bra}[1]{ \langle {#1} | }
\nc{\ket}[1]{ | {#1} \rangle }
\nc{\fb}{\,{\rm fb}^{-1}}
\nc{\ev}{{\rm eV}}
\nc{\kev}{{\rm keV}}
\nc{\Mev}{{\rm MeV}}
\nc{\gev}{{\rm GeV}}
\nc{\tev}{{\rm TeV}}
\nc{\mev}{{\rm MeV}}

\def\del{\partial}
\def\Dslash{\not{\hbox{\kern-4pt $D$}}}
\def\dslash{\not{\hbox{\kern-2pt $\del$}}}
\def\pslash{\not{\hbox{\kern-2pt $p$}}}
\def\ETmiss{ \not{\hbox{\kern-4pt $E$}}_T }

\def\msb{{\bar{\ssstyle M \kern -1pt S}}}

\def\babar{\mbox{\sl B\hspace{-0.4em} {\small\sl A}\hspace{-0.37em} \sl B\hspace{-0.4em} {\small\sl A\hspace{-0.02em}R}}}


%% file: introduction_arx.tex
\section{Introduction}
\label{sec:intro}

\begin{figure}[htbp]
\begin{center}
\includegraphics[width=5cm]{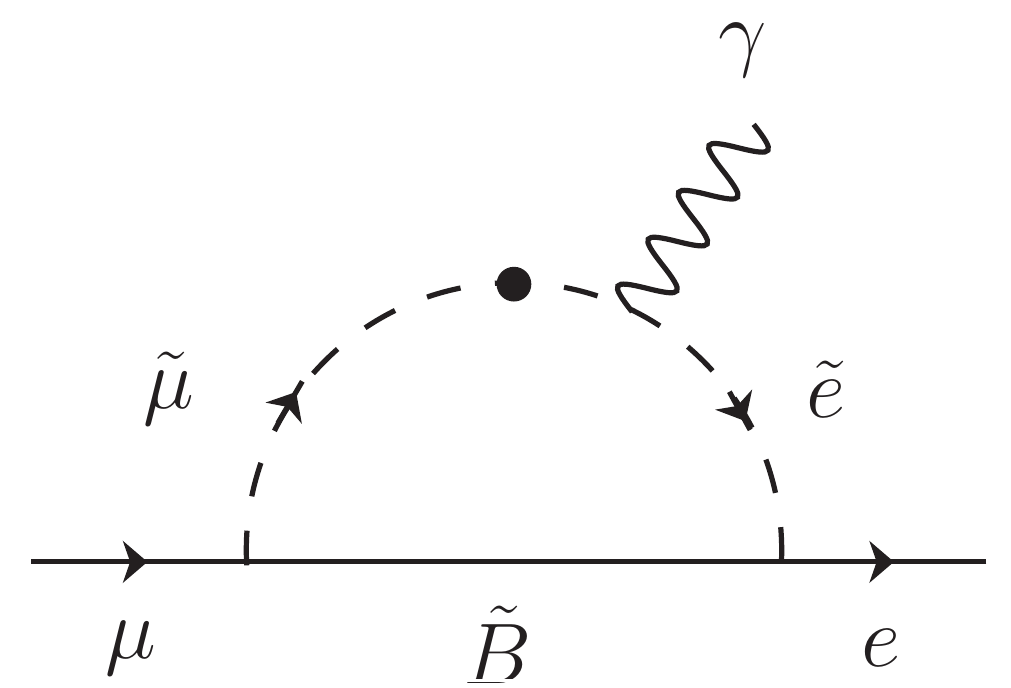} \hspace*{2cm}
\includegraphics[width=5cm]{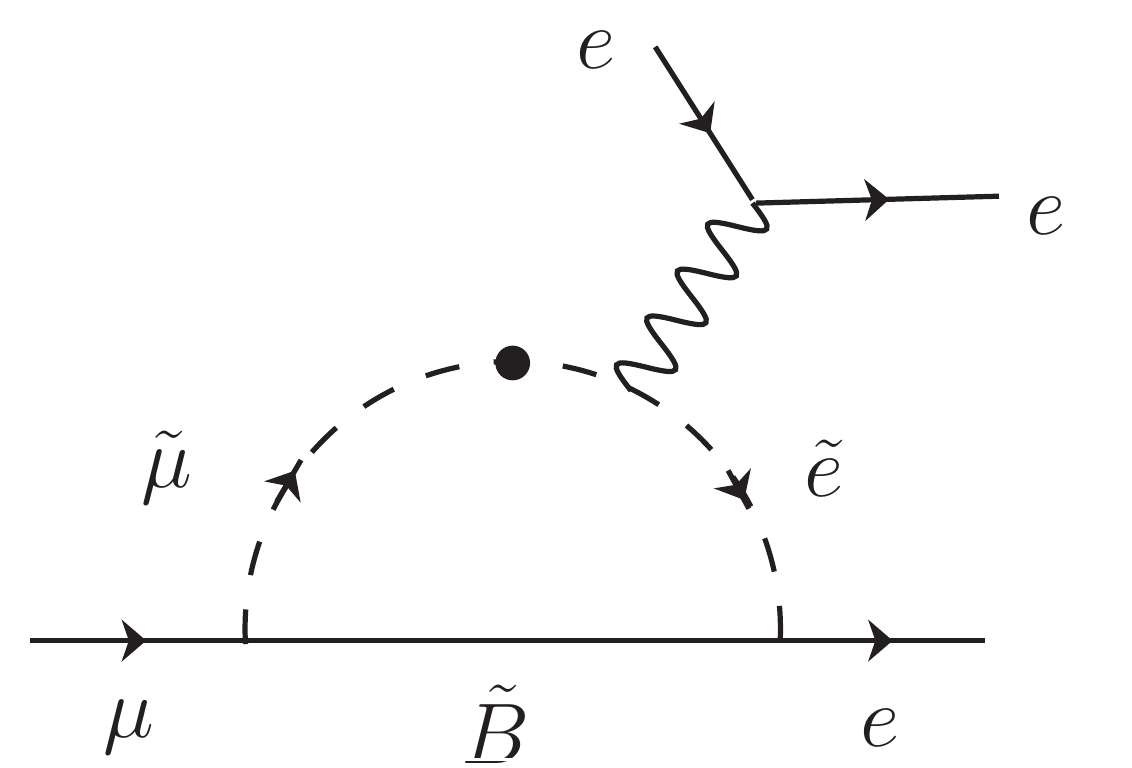}
\caption{\label{CL:mutoegamma}$\mu \rightarrow e\gamma$ decay mediated by SUSY particles (left panel), and $\mu \rightarrow 3e$ decay (right panel).}
\end{center}
\end{figure}

Charged lepton flavor violating (CLFV) processes, such as 
$\mu^+\to e^+\gamma$ and $\mu^+\to e^+e^-e^+$, are mediated by neutrino oscillations in loop diagrams in the Standard Model (SM). While allowed, these reactions are highly suppressed due to the extremely small neutrino masses. For example, the branching fraction for $\mu \to e \gamma$ is given by
\begin{equation}
BR(\mu \rightarrow e \gamma)=\frac{3\alpha}{32\pi}\left|\sum_i U_{\mu
i}^* U_{e i}\frac{m_{\nu_{i}}^2}{m_{W}^2}\right|^2 \sim 10^{-52},
\end{equation}
where $U_{ei}$ are the leptonic mixing matrix elements, assuming neutrinos
are Dirac particles. This is clearly well below the reach of any conceivable experiment.
However, in many extensions of the SM, such as supersymmetric grand unified
theories or theories with extra dimentions, larger contributions to CLFV are
allowed~\cite{new physics}. Observing CLFV is therefore a clear indication of physics beyond the Standard Model (henceforth BSM physics). Figure~\ref{CL:mutoegamma} shows an example of BSM processes mediated by SUSY particles.

The effective Lagrangian relevant for the $\mu^+\to e^+\gamma$ and $\mu^+\to e^+e^-e^+$
decays can be parametrized, regardless of the origin of CLFV, as a sum of
dipole terms and a ``contact term''. The $\mu^+\to e^+\gamma$ process is only sensitive to the dipole terms, while both dipole and contact terms contribute to $\mu^+\to e^+e^-e^+$ decays~\cite{deGouvea:2013zba}. Improving upper limits of
the $\mu^+\to e^+\gamma$ and $\mu^+\to e^+e^-e^+$ branching fractions down to $10^{-14}$ and $10^{-16}$, respectively, could probe scales of BSM physics up to several thousands of TeV. In addition, the Dalitz plot of the $\mu \rightarrow eee$ decays offers the possibility to determine the chirality of BSM physics, should it be observed with sufficient statistics~\cite{Okada:1999zk}. 

We discuss herein feasibility studies of next generation detectors designed to search for $\mu^+\to e^+\gamma$ and $\mu^+\to e^+e^-e^+$ decay that could be performed at Fermilab during Project X era. These searches complement improved searches for $\mu \to e$ conversion that could also be done at Project X~\cite{Mu2eII}.

%% file: mutoegamma_arx.tex
\section{$\mu^+\to e^+\gamma$}

Recent MEG measurement at PSI~\cite{MEG2013} sets a limit of 
${\cal B}(\mu^+\to e^+\gamma)< 5.7\times 10^{-13}$ at 90\% confidence
level using $3.6\times 10^{14}$ stopped muons on target. The MEG detector
consists of a set of drift chambers and scintillation timing counters,
located inside a superconducting solenoid, and a liquid Xenon 
calorimeter with UV-sensitive photomultiplier tubes, located outside the
solenoid. 

There are two main sources of background. Over 90\% of the background in
the signal region comes from accidental background, that is, a positron
from a regular Michel muon decay combined with a photon from a radiative
muon decay  (RMD) $\mu^+\to e^+ \nu_e\bar\nu_\mu \gamma$.
Most of the remainig background is due to RMD where the neutrinos carry away minimum
energy. The accidental background rate depends on the instantaneous stopping
muon rate $R_\mu$, total integrating data acquisition time $T$, and 
detector resolutions:
\begin{equation}
N_{\rm acc} \propto R_\mu^2 \times \Delta E_\gamma^2 \times\Delta P_e \times
\Delta \Theta_{e\gamma}^2 \times \Delta t_{e \gamma} \times T, 
\end{equation}
where $\Delta E_\gamma$ and $\Delta P_e$ are the resolutions of photon energy
and positron momentum, respectively; $\Delta \Theta_{e\gamma}$ and
$\Delta t_{e \gamma}$ are the resolutions of $e\gamma$ opening angle and
timing.

The MEG Collaboration has proposed an upgrade~\cite{MEGupgrade} aiming to 
improve the sensitivity to $\mu\to e\gamma$ decay 
by one order of magnitude below the current limit, {\it i.e.,} to set a limit at
$\sim 6\times 10^{-14}$ in the absence of signal. They will replace 
their tracker with a lower-mass, higher-granularity device, reduce target
thickness, use a faster timing counter array, and increase the 
granularity of the liquid xenon detector by replacing the PMTs with 
a larger number of smaller solid state photosensors. The sensitivity
estimate is based on a muon stopping rate of $7\times 10^7$ muons/s for 
a three year run, assuming 180 DAQ days per year.

To improve the experimental reach beyond that of the MEG upgrade, one needs to further improve the detector 
sensitivity. 
The photon energy resolution is a major limiting factor in this search.
A pair spectrometer that measures $e^+e^-$ pair tracks from photon
conversions in a thin dense material can greatly improve the photon energy
resolution. This approach was discussed at  2012 Project X Summer Study~\cite{Fritz}. The loss of efficiency due to the small photon conversion 
probability can be compensated for by improved fiducial solid angle coverage and by the higher beam power at Project X at 
Fermilab. 

We have conducted an initial study of this concept using a fast simulation
tool (FastSim) originally developed for the Super$B$ experiment~\cite{SuperB}
using the \babar\
software framework and analysis tools. FastSim allows us to model detector 
components as two-dimensional shells of simple geometries such as cylinders,
cones, disks, and planes. The effect of physical thickness is modeled
parameterically. Coulomb scattering and ionization energy loss are modeled
with the standard parameterization in terms of radiation length and particle
momentum and velocity. Bremsstrahlung and pair production are modeled by
simplified cross-sections. Tracking measurements are described in terms of
the single-hit and two-hit resolution, and the efficiency. Silicon strip
detectors are modeled as two independent orthogonal projections. 
FastSim reconstructs high-level detector objects from simulated hits and
energy deposits using the simulation truth to associate detector objeccts,
bypassing pattern recognition. Errors associated with pattern recognition are
introduced by perturbing the truth-based association, using models based on
\babar\ pattern recognition algorithm performance. The final set of hits on
associated with a track is passed to the \babar\ Kalman filter track fitting
algorithm to obtain reconstructed track parameters.

\begin{figure}[htbp]
\centering
\begin{minipage}[c]{0.47\textwidth}
\centering
\includegraphics[width=\textwidth]{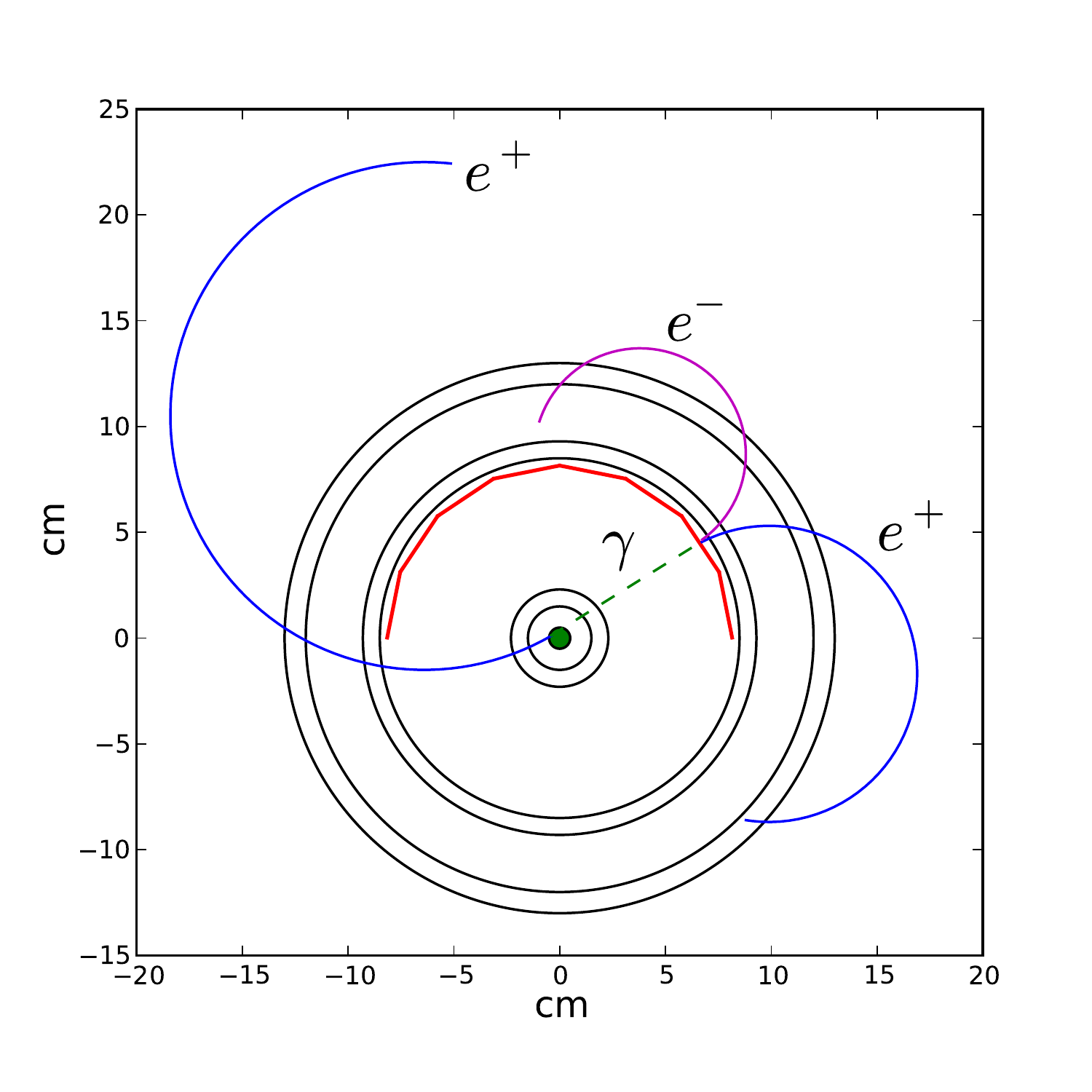}
\caption{Schematic drawing (in the plane transverse to the muon beam axis) of the $\mu\to e\gamma$ detector.}
\label{fig:detscheme}
\end{minipage}
\quad
\begin{minipage}[c]{0.47\textwidth}
\centering
\includegraphics[width=\textwidth]{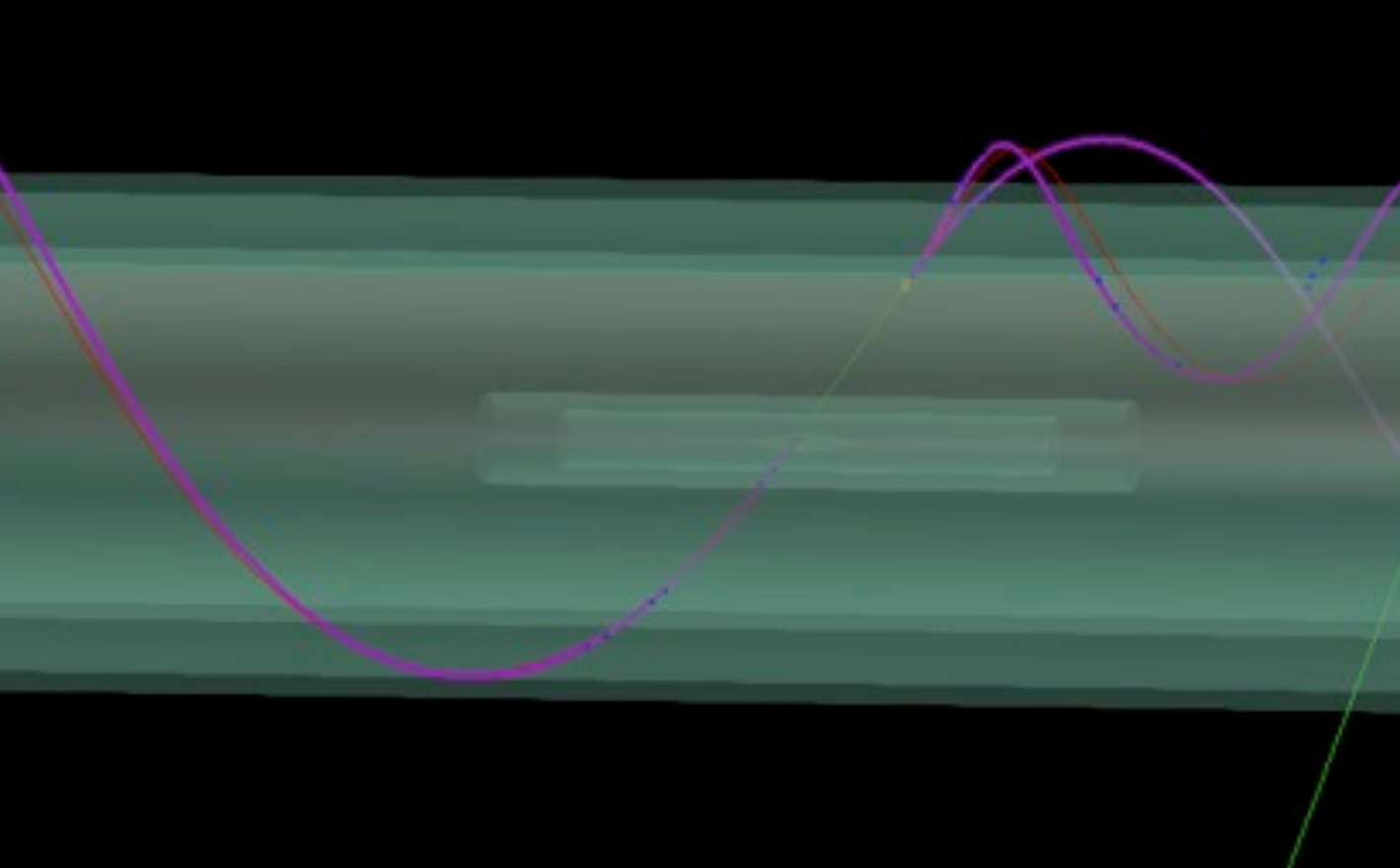}
\caption{FastSim signal event display}
\label{fig:evtdisplay}
\end{minipage}
\end{figure}

The FastSim model in this study consists of a thin aluminum stopping target 
and a six-layer
cylindrical silicon detector. A 0.56 mm thick lead (10\% $X_0$) half cylinder 
covering 0--$\pi$ in azimuthal angle at $R = 80$ mm serves as the photon converter.
The target consists of two cones connected at their base; each cone is 50 mm 
long, 5 mm in radius, and 50 $\mu$m thick. Two silicon detector cylinders are
placed close the target for better vertexing resolution; two layers are placed
just outside the Pb converter, and two layers a few cm away. The layout is shown
in Fig.\ref{fig:detscheme}; a signal event display is shown in 
Fig.~\ref{fig:evtdisplay}. The silicon detector is modeled after Super$B$ 
inner silicon striplet modules but thinner. Each layer is formed of 50 $\mu$m thick 
double-sided striplets silicon sensors mounted on 50 $\mu$m of kapton. 
The hit spatial resolution is modeled as a sum of two components with resolutions
of 8 $\mu$m and 20 $\mu$m, and a hit efficiency of 90\%.
The entire detector is placed in a 1T solenoidal
magnetic field.

\begin{figure}[ht]
\centering
\includegraphics[width=0.49\textwidth]{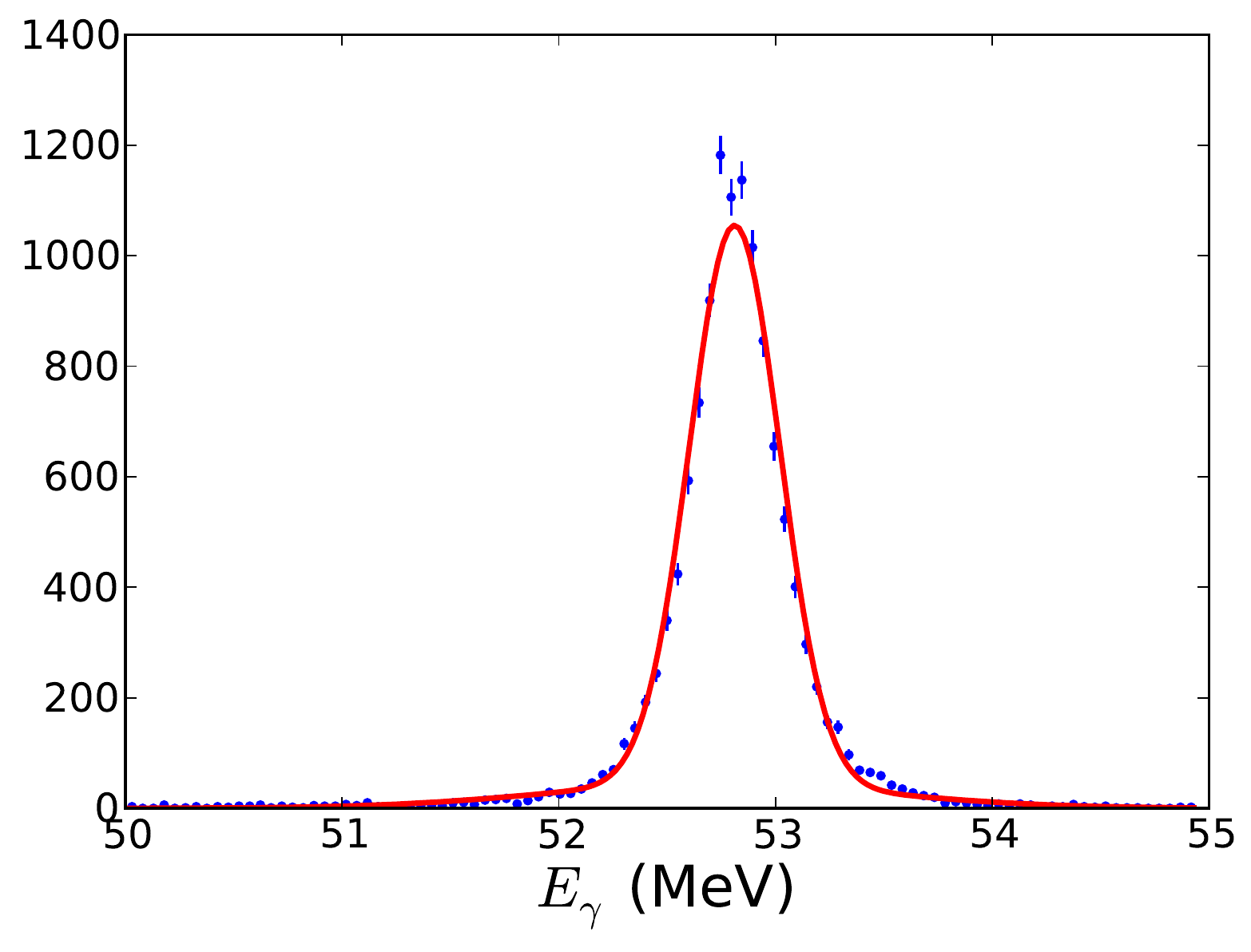}
\includegraphics[width=0.49\textwidth]{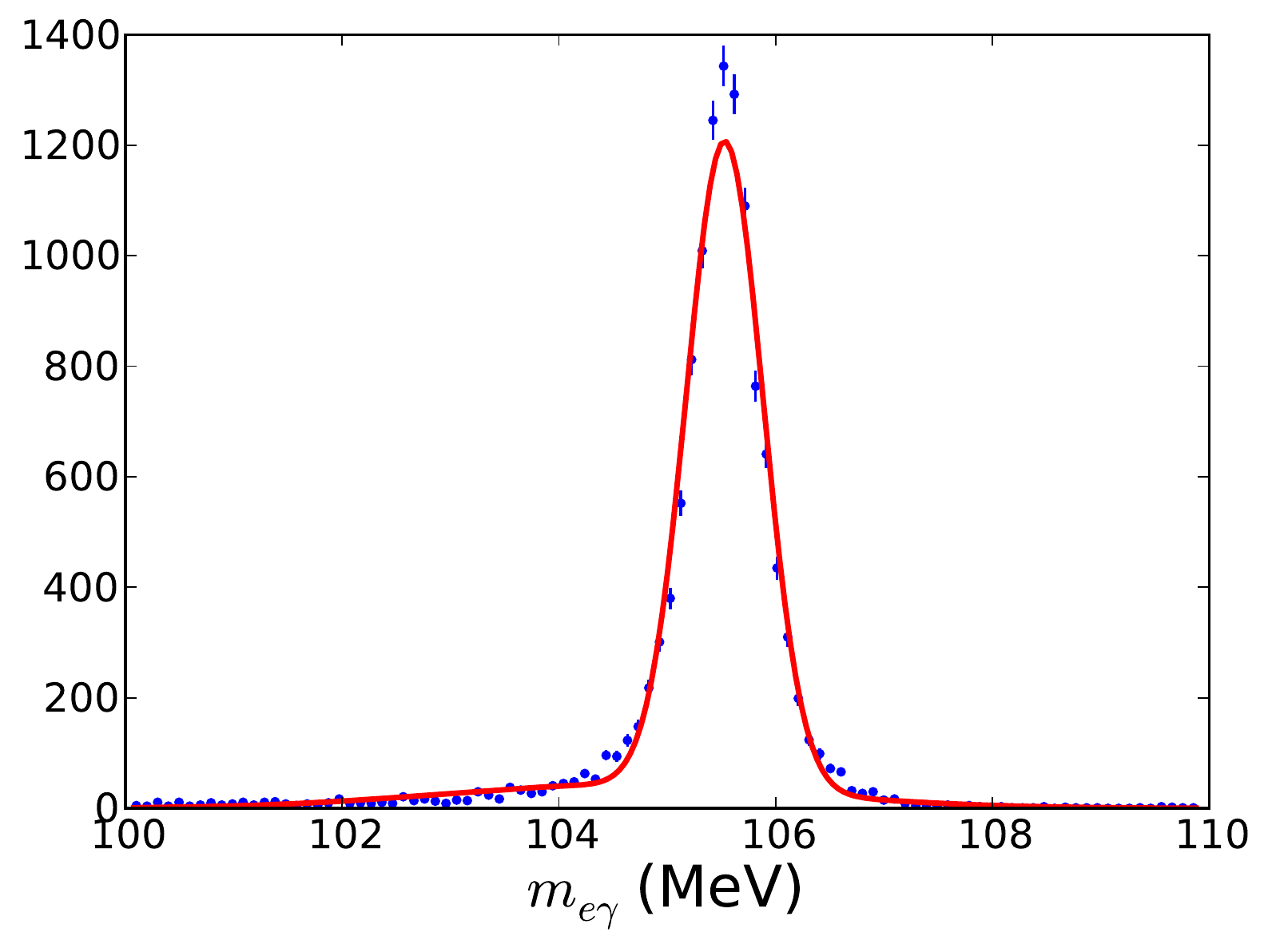}
\caption{\label{fig:eresol} Photon energy and $e\gamma$ invariant
mass distributions. Fitted curve is a double-Gaussian distribution.}
\end{figure}

We generate muons at rest and have them decay via $\mu^+\to e^+\gamma$
to study the reconstruction efficiency and resolution. 
Approximately 1.3\% of generated signal events are well-reconstructed, 
passing quality and fiducial selection criteria. The photon energy resolution 
is approximately 200~keV (Fig.~\ref{fig:eresol}), similar to the positron momentum
resolution, which 
corresponds to 0.37\% for 52.8 MeV photons. This is a substanial improvement compared 
to the 1.7\%--2.4\% resolution of the current MEG and the 1.0\%--1.1\% resolution 
goal of the MEG upgrade. 

The positron angular resolution is slightly below 10~mrad in both $\theta$ 
and $\phi$ views,
better than current MEG performance but worse than MEG upgrade projection.
The photon direction, determined solely from $e^+e^-$ momenta, has a resolution
similar to that of the positron. It can be further improved  by using
vertex information. Both the  $\gamma\to e^+e^-$ vertex and positron production vertex
(by extrapolating positron track back to the target) have a position resolution
of the order of 100~$\mu$m. Therefore, the photon direction, determined by connecting the two
vertices, has a resolution of the order of 1~mrad (given the lever arm of 80 mm).
As a result, the resolution of the angle between $e^+$ and $\gamma$ is dominated
by $e^+$ angular resolution.

We then use a toy Monte Carlo technique to determine the sensitivity of this
apparatus.
For accidental background, we generate $e^+$ and $\gamma$ from the Michel
spectrum and the RMD spectrum~\cite{Kuno:1999jp}, respectively. Only those momenta
near the end points of the spectra could contribute to the background.
The directions, production points, and production times of $e^+$ and 
$\gamma$ are generated
randomly without correlation. We ignore the other positron originating from the RMD.
For the RMD background, we generate $e^+$ and $\gamma$ according to the theoretical
partial branching fraction formula~\cite{Kuno:1999jp}. Their directions are
correlated, and their production times and positions are identical.
The number of accidental background events is a product of $R_\mu^2$, the partial
branching fractions of the Michel decay and RMD, the selection timing window, the
total DAQ time, phase space factors, and the reconstruction and selection 
efficiencies. For the RMD background, the scaling factor is $R_\mu$, instead of
$R_\mu^2$.

\begin{figure}[htbp]
\includegraphics[width=0.99\textwidth]{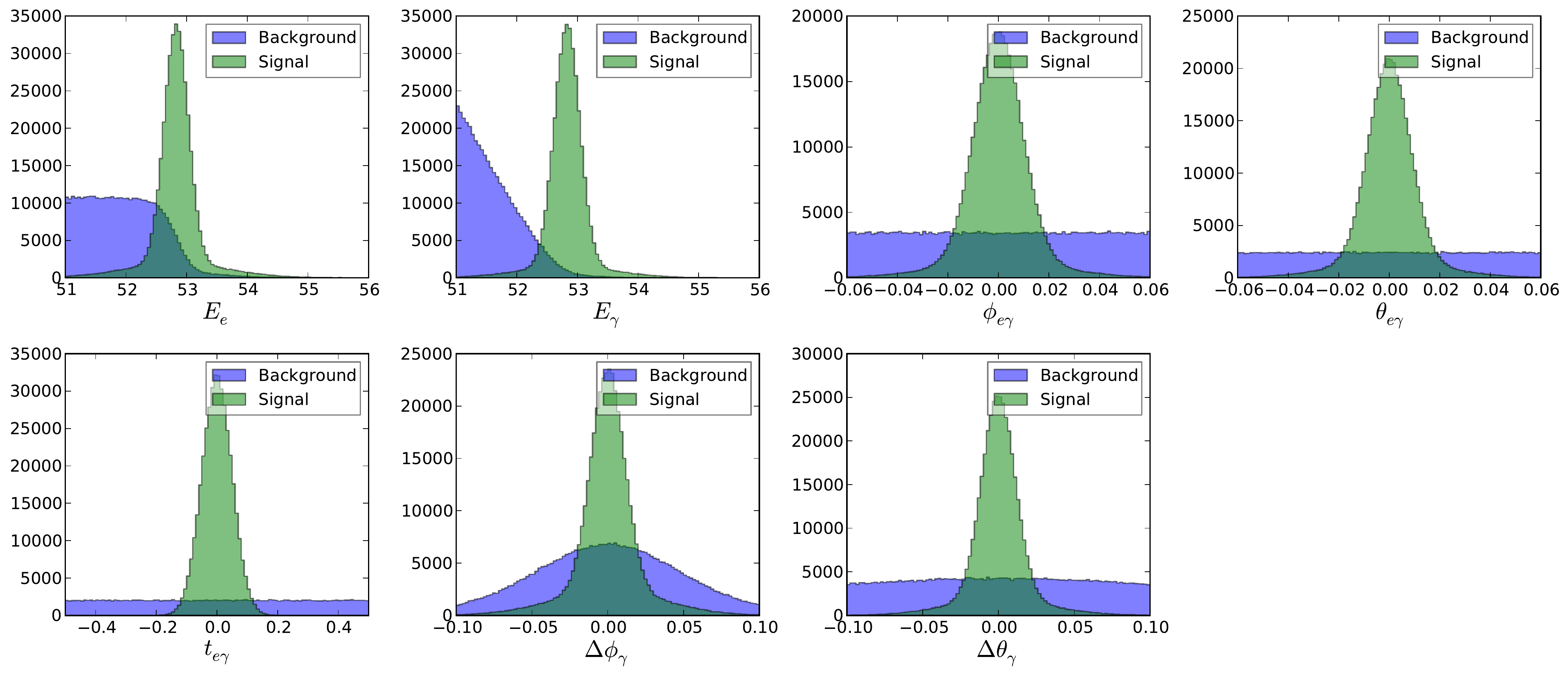}
\caption{Discriminating variables used in the $\mu^+\to e^+\gamma$ search.}
\label{fig:muegamma-vars}
\end{figure}

The energies and directions
of the $e^+$ and $\gamma$ are smeared according to the FastSim study
using double-Gaussian functions.
We study the scenarios with timing resolutions of 50~ps and 100~ps. 
The MEG experiment uses 5 independent variables $E_\gamma$, $p_e$, 
$\phi_{e\gamma}$, $\theta_{e\gamma}$, and $\Delta t_{e\gamma}$, to construct
their likelihood function. In our detector, we can take advantage of the excellent
direction resolution of the converted photon. If the photon is produced
at a different point from positron production point, as is the case for accidental backgrounds,
the direction of the $\gamma\to e^+e^-$ momentum and that of the line
connecting the $e^+e^-$ vertex and the primary $e^+$ production point on the target
will be different.  Two additional variables $\Delta\theta_\gamma$ 
and $\Delta\phi_\gamma$ are therefore used in our study. Comparisons between signal
and accidental background are shown in Fig.~\ref{fig:muegamma-vars}.

To estimate the 90\% C.L. upper limit sensitivity, we use a cut-and-count
approach to estimate the background level and then a Feldman-Cousins 
method~\cite{Feldman:1997qc} to calculate the upper limit sensitivity assuming
no signal events are present.

\begin{figure}[htbp]
   \centering
   \includegraphics[width=0.48\textwidth]{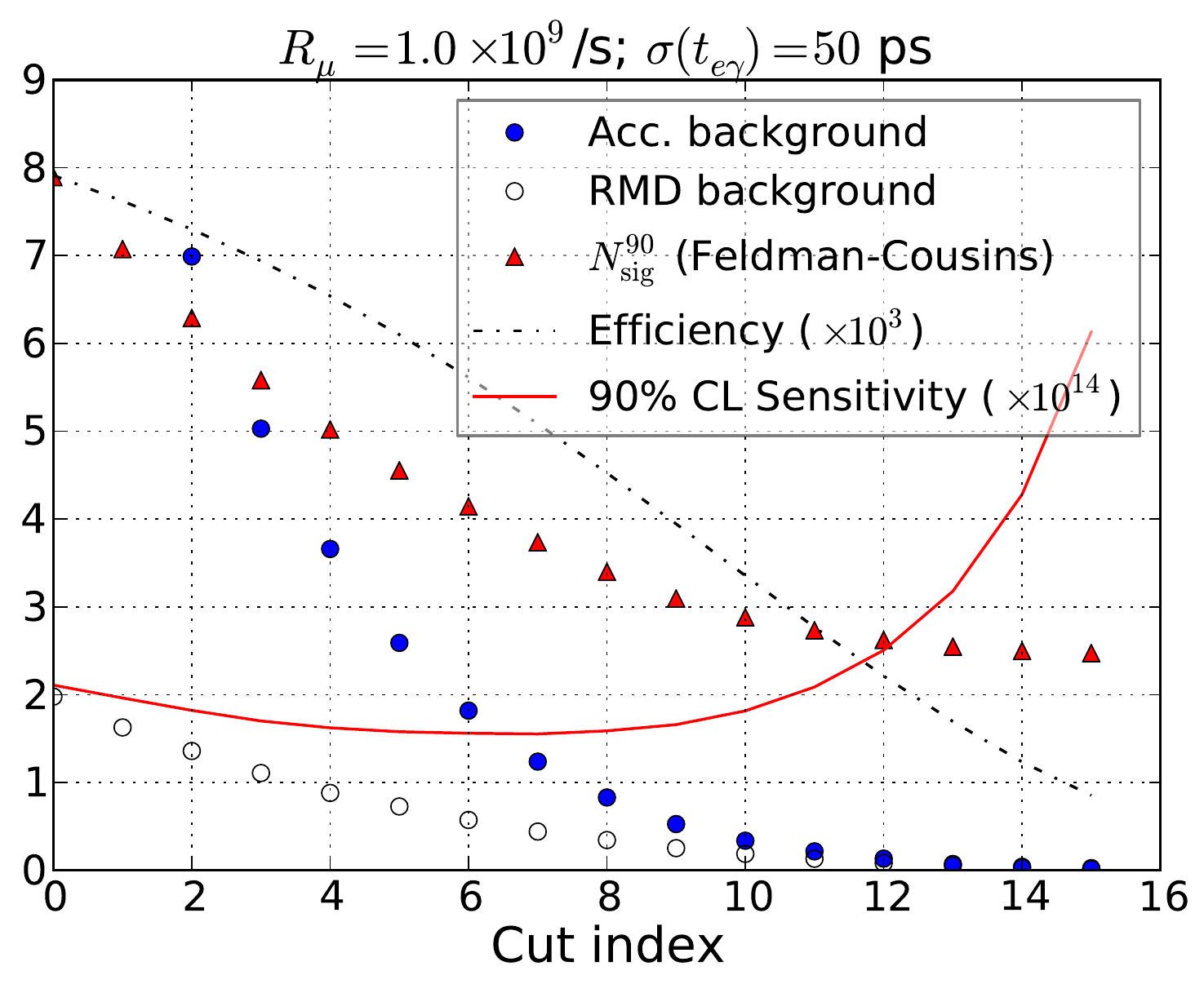} 
   \includegraphics[width=0.48\textwidth]{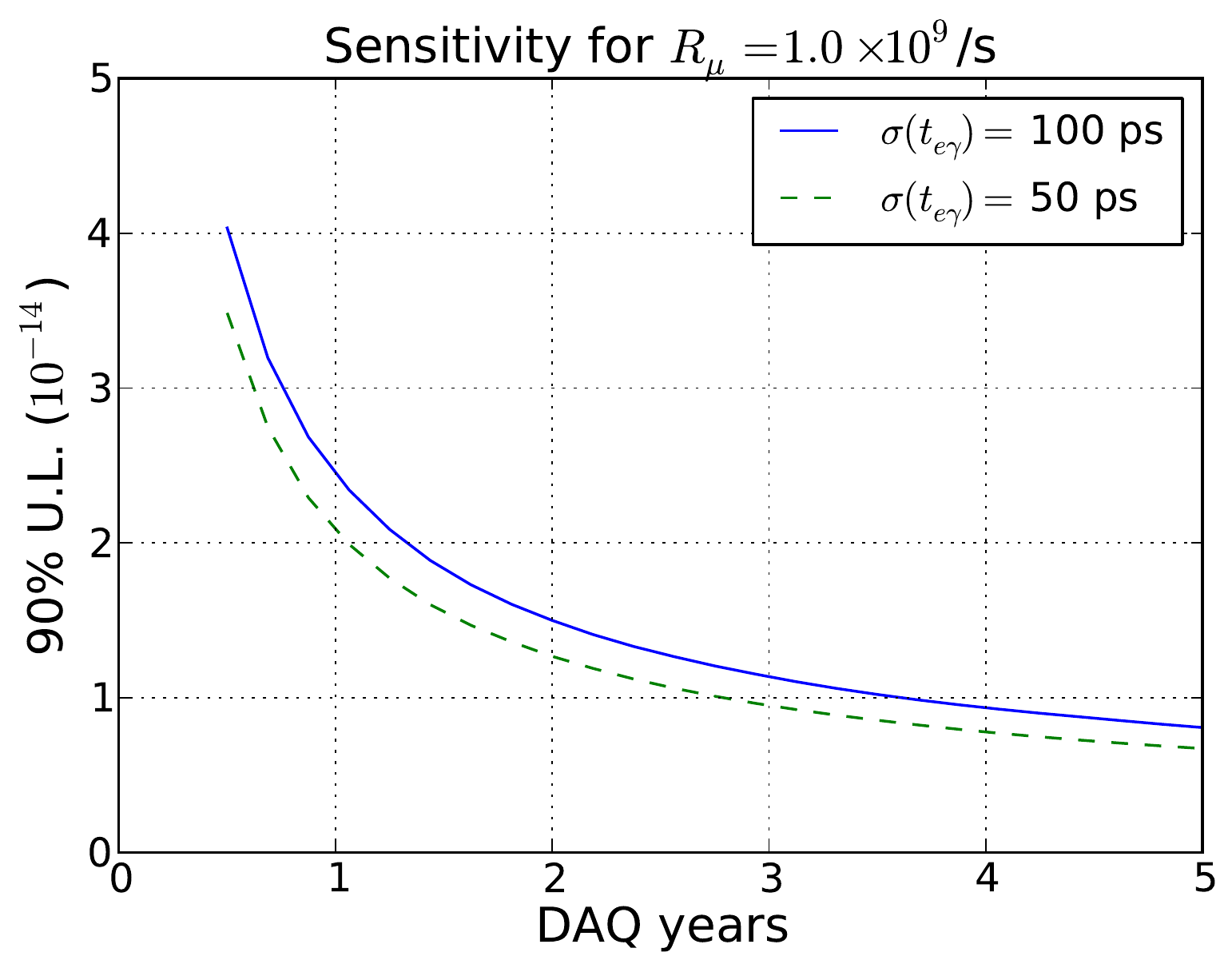} 
   \caption{Left: $B(\mu^+\to e^+\gamma)$ sensitivity optimzation for a given
scenario (see text). Right: sensitivity as a function of integrated DAQ time for both 50-ps and 100-ps $t_{e\gamma}$ resolutions.}
   \label{fig:muegamma-sensitivity}
\end{figure}

Figure~\ref{fig:muegamma-sensitivity} shows the background levels, 
signal efficiency, and 90\%
C.L. sensitivity under various selection cuts for 
$R_\mu=1\times 10^{9}$~muons/s, and 50-ps resolution on $t_{e\gamma}$.
A sensitivity of $B(\mu^+\to e^+\gamma)<1.6\times 10^{-14}$ could be reached
with an integrated DAQ time of 1.5 years.
The sensitivity reach
as a function of integrated DAQ time for both 50-ps and 100-ps timing
resolutions is also shown.

\begin{figure}[htbp]
\centering
\includegraphics[width=0.48\textwidth]{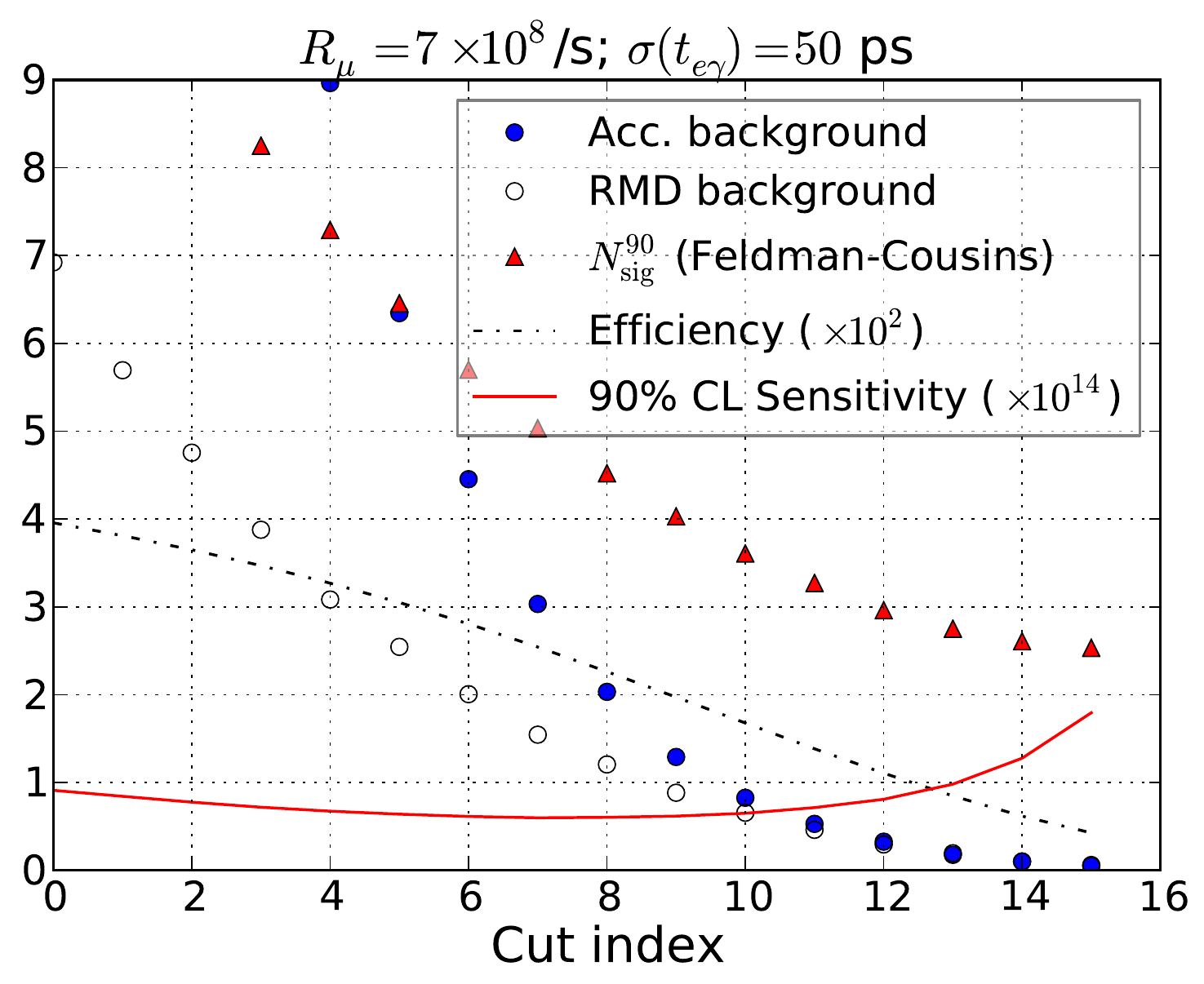}
\caption{Left: $B(\mu^+\to e^+\gamma)$ sensitivity optimzation with 5-times
higher signal sensitivity and lower $R_\mu$ than that in 
Fig.~\ref{fig:muegamma-sensitivity}. Best sensitivity is $6\times 10^{-15}$.}
\label{fig:muegamma-sens-5x}
\end{figure}

Increasing the muon rate further could improve the sensitivity. However,
the sensitivity quickly moves away from the ${\cal O}(1)$ background regime, because the accidental
background grows as $\sim R_\mu^2$. A better approach is to increase the
efficiency and reduce the muon rate to keep the background level low. 
Figure~\ref{fig:muegamma-sens-5x} 
shows a scenario in which the signal efficiency is 5-times higher
and the muon stopping rate is slightly reduced to $R_\mu=7\times 10^{8}$. 
In this scenario, one can reach a sensitivity of $B(\mu^+\to e^+\gamma)<6\times 10^{-15}$.
Such an approach can be realized with multiple layers of thin photon converters and associated silicon tracking layers. Studies of the sensitivity of a multi-converter design are underway.

An alternative version of the photon conversion approach to a $\mu \to e \gamma$ 
experiment has also been discussed~\cite{franco}. In this version, consider a large volume
solendoidal magnet, such as the KLOE coil, which has a radius of 2.9~m, run at a
field of perhaps 0.25~T. A large volume, low mass cylindrical drift chamber 
provides many ($\ge$100) layers of tracking, utilizing small cells and having 
a total number of sense wires approaching $10^5$. Interspersed every ten layers 
is a 0.5 mm W converter shell. There are a sufficient number of points on the 
$e^+$ and $e^-$ tracks from converted photons behind each converter to reach a
 total conversion efficiency of perhaps 80\%, with excellent photon mass 
resolution.

In summary, using a converted photon to increase the $\mu^+\to e^+\gamma$ detection
sensitivity by improving the photon energy resolution appears to be a promising approach. More detailed studies are 
needed to quantify the requirements in detail, with the goal of improving upon the MEG upgrade
sensitivity by an about order of magnitude.

%% file: mutoeee_arx.tex
\section{$\mu^+\to e^+e^-e^+$}
\label{mutoeee}

The current bound on the $\mu^+\to e^+e^-e^+$ decay has been set by the SINDRUM experiment at PSI~\cite{Bellgardt:1987du}. No signal was observed; a limit of ${\cal B}(\mu^+\to e^+e^-e^+) < 1\times 10^{-12}$ was therefore derived, assuming a decay model with a constant matrix element. This measurement was limited by the number of stopped muons, the background from $\mu^+\to e^+e^-e^+ \nu_e \bar\nu_\mu$ decays remaining negligible. The Mu3e experiment~\cite{Blondel:2013ia} has been proposed to improve this bound by four orders of magnitude, reaching a single event sensitivity (SES) at the level of $7 \times 10^{-17}$. The experiment consists of a silicon pixel detector immersed in a ~1 T magnetic field and surrounding a double-cone target, and two timing detector systems. The dominant backgrounds arise from $\mu^+ \rightarrow e^+e^-e^+ \overline{\nu}_{\mu} \nu_e$ events, as well as accidental coincidences of tracks from $\mu^+ \rightarrow e^+ \overline{\nu}_{\mu} \nu_e$ and $\mu^+ \rightarrow e^+e^-e^+ \overline{\nu}_{\mu} \nu_e$ decays. Excellent momentum resolution ($<0.5 \Mev$) and timing resolution (50-500~ps depending on the detector system) reduce these backgrounds at an acceptable level.

\begin{figure}[htbp]
\minipage{0.35\textwidth}
\begin{center}
\includegraphics[width=\textwidth]{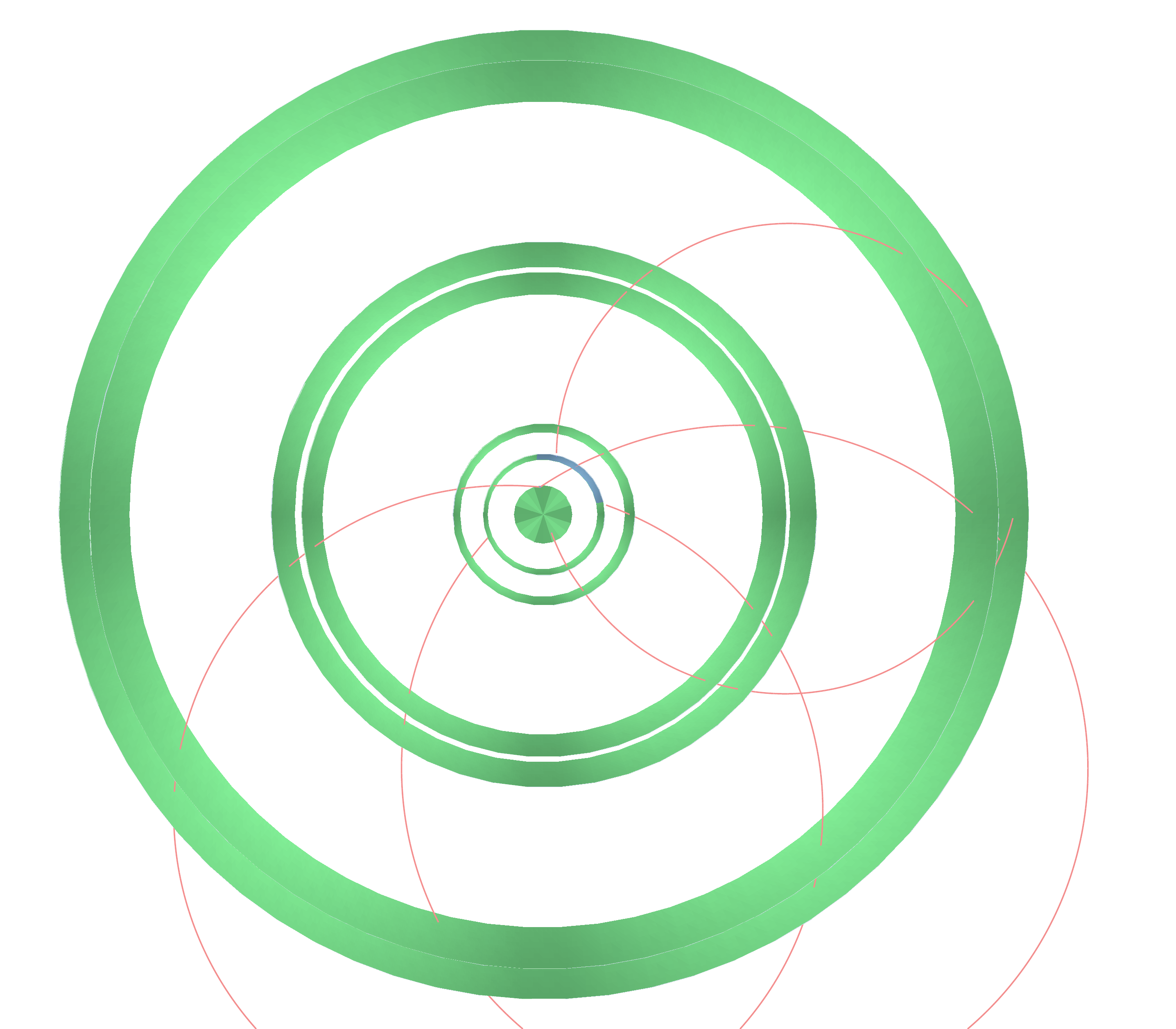}
\end{center}
\endminipage\hfill
\minipage{0.55\textwidth}
\begin{center}
\includegraphics[width=\textwidth]{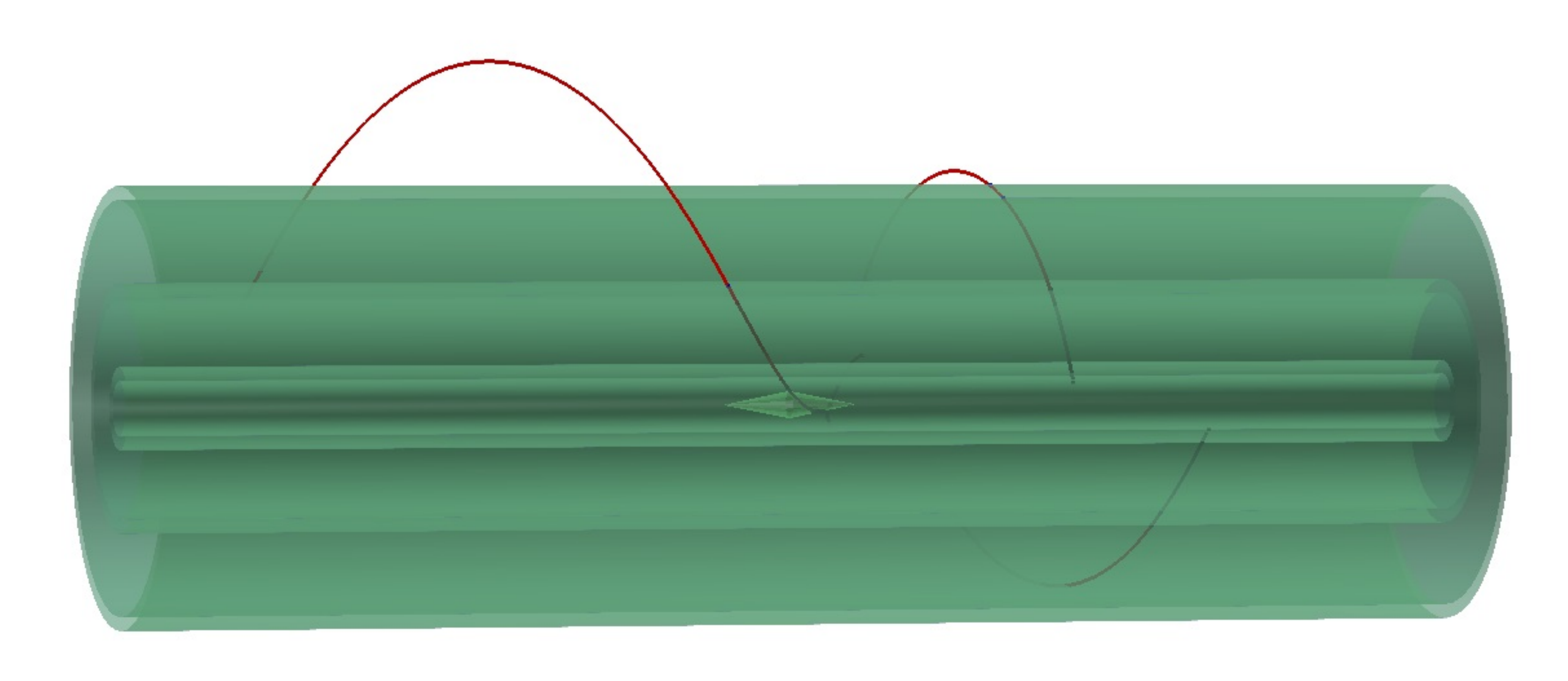}
\end{center}
\vspace{0.5cm}
\endminipage\hfill
\caption{Display of the experimental setup, together with a simulated $\mu^+ \rightarrow e^+e^-e^+$ event.}
\label{Fig::mu3e}
\end{figure}

Our study aims to increase the expected Mu3e sensitivity by an order of magnitude. This requires an improved detector to further reduce the physics and accidental backgrounds. We employed the fast simulation tool discussed above, and explored the improvements needed to achieve a SES at the level of $5 \times 10^{-18}$. The FastSim model consists of a silicon tracker composed of 6 cylindrical layers, surrounding an active target. Each layer is formed of 50 $\mu$m thick double-sided striplet silicon sensors mounted on 50 $\mu$m of kapton. The hit spatial resolution is modeled as a sum of two components with resolutions of 8 $\mu$m and 20 $\mu$m, and a hit efficiency of 90\%. The active target is made of two hollow cones of silicon pixel detectors connected at their base. Each cone is 5~cm long, 50 $\mu$m thick and has a radius of 1~cm, with a pixel size of 50 $\mu$m by 50 $\mu$m. Although not included in the simulation, a time-of-flight system should be installed as well. We assume a time resolution of 250 ps, averaging the values of the corresponding Mu3e detector systems. The apparatus layout is displayed in Fig.~\ref{Fig::mu3e}, together with a simulated $\mu^+ \rightarrow e^+e^-e^+$ event.

\begin{figure}[htb]
\begin{center}
\includegraphics[width=0.6\textwidth]{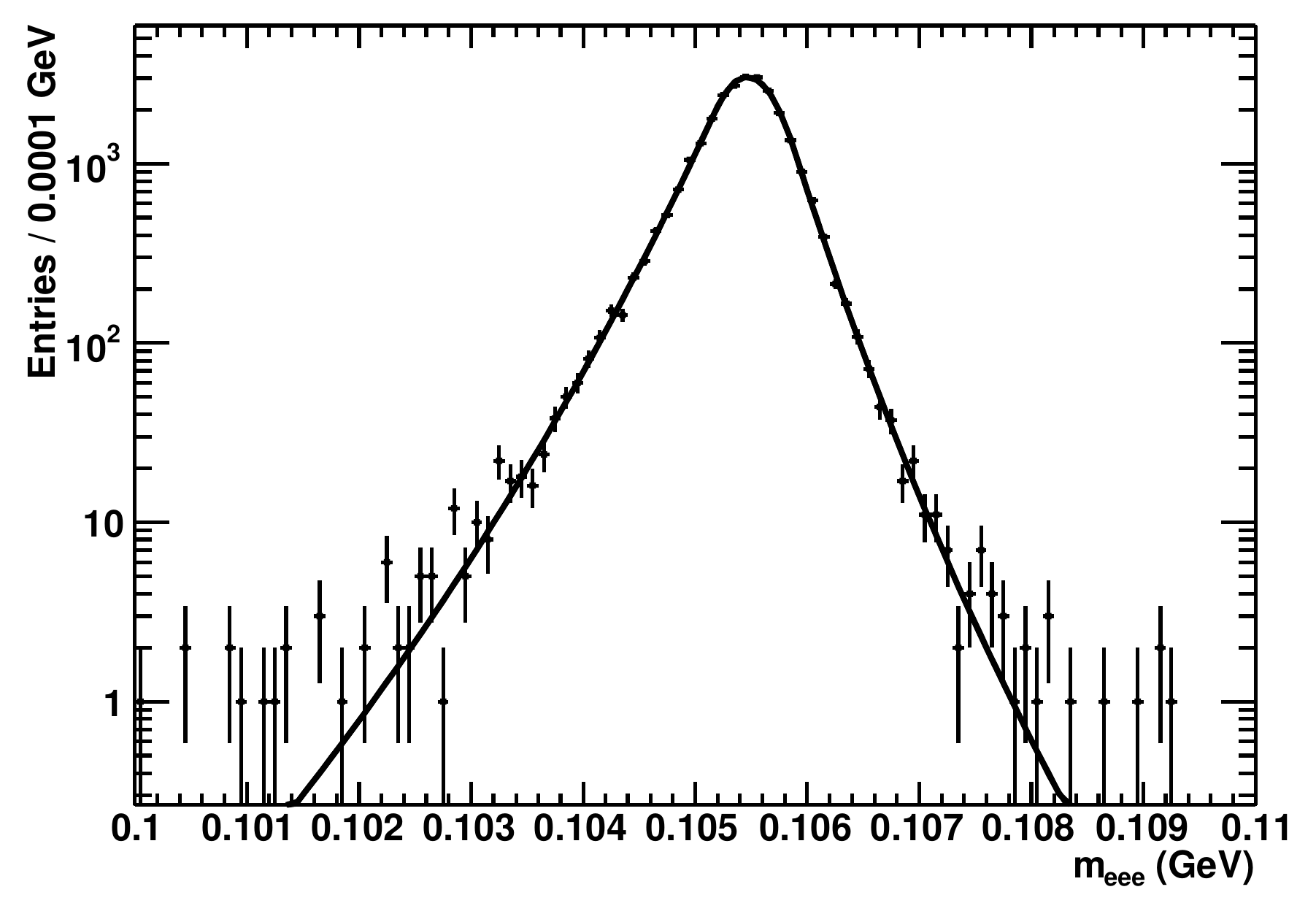}
\end{center}
\caption{The $e^+e^-e^+$ invariant mass distribution after all selection criteria are applied fitted by a double-sided Crystal Ball function.}
\label{Fig::mu3e2}
\end{figure}

We generate $\mu^+ \rightarrow e^+e^-e^+$ events according to phase space to study the detector resolution and efficiency. The stopped muons are reconstructed by combining three electrons, constraining the tracks to originate from the same pixel in the active target. To further improve the resolution, we require the probability of the constrained fit to be greater than 1\%, and a reconstructed muon momentum less than $1 ~\Mev$. The absolute value of the cosine of the polar angle of each electron must also be less than 0.9. The resulting $e^+e^-e^+$ invariant mass distribution, shown in Fig.~\ref{Fig::mu3e2}, peaks sharply at the muon mass. We extract the resolution by fitting this spectrum with a double-sided Crystal Ball function (a Gaussian with power-law tails on both sides). The Gaussian resolution is found to be $0.3~\Mev$. To investigate the contribution of the active target to the resolution, we performed alternative fits, removing the geometric constraints, or taking the vertex position by considering all points from tracks intercepting the target, and choosing the one minimizing the $\chi^2$ of the constrained fit. While we observe an improvement compared to the unconstrained fit, the second method yields a similar signal resolution. However, the active target provides a better estimate of impact parameters of the tracks, improving background rejection.

The signal efficiency is found to be 27\%. To achieve a SES at the level of $5\times\sim 10^{-18}$ after a 3-year run with 100\% DAQ efficiency, a stopped muon rate of the order of $8\times 10^{9}$ is needed. For comparison, the Mu3e stopped muon rate at the HiMB beam at PSI is expected to be of the order of $2\times 10^{9}$.   

\begin{figure}[htb]
\begin{center}
\includegraphics[width=0.45\textwidth]{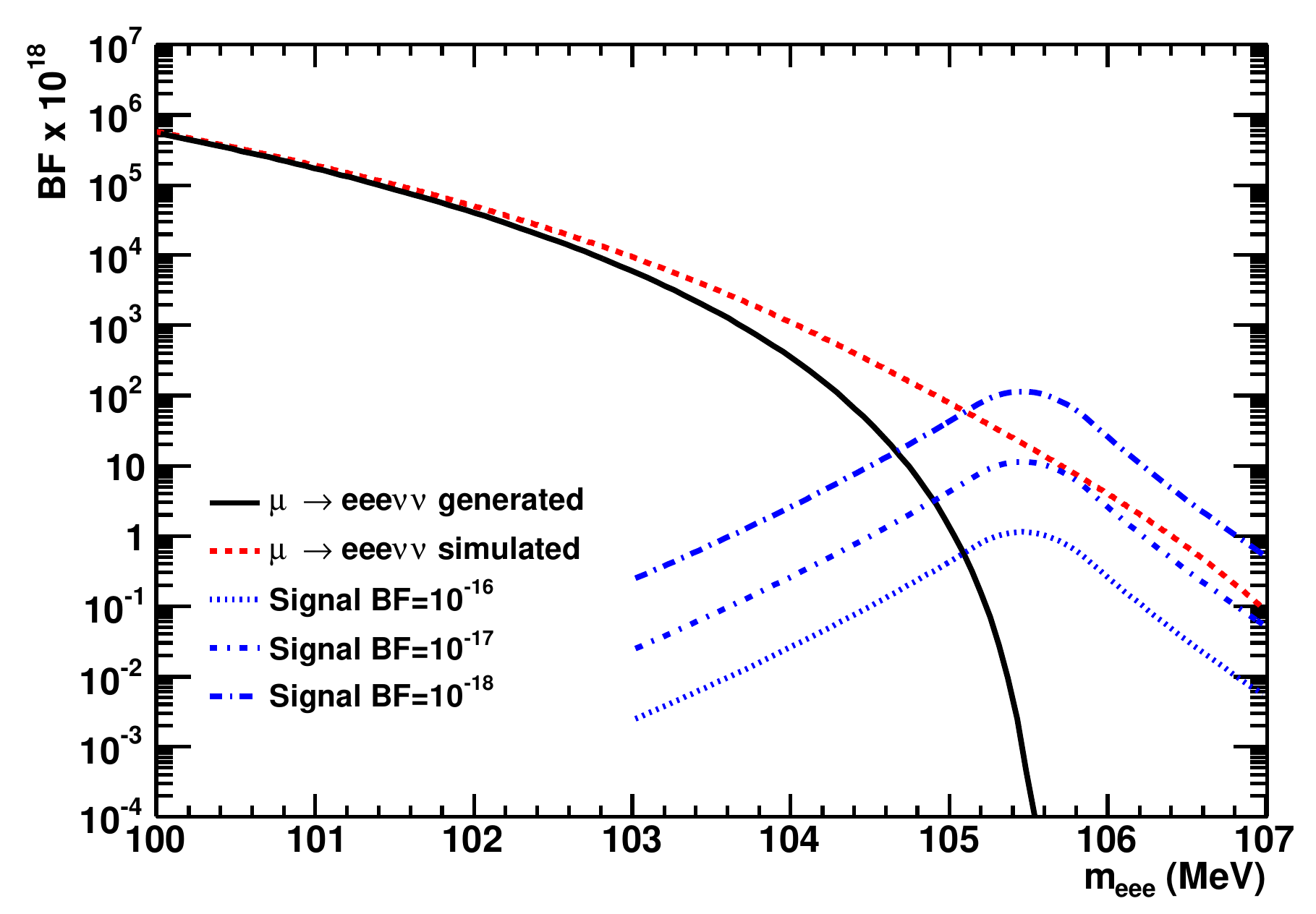}
\includegraphics[width=0.45\textwidth]{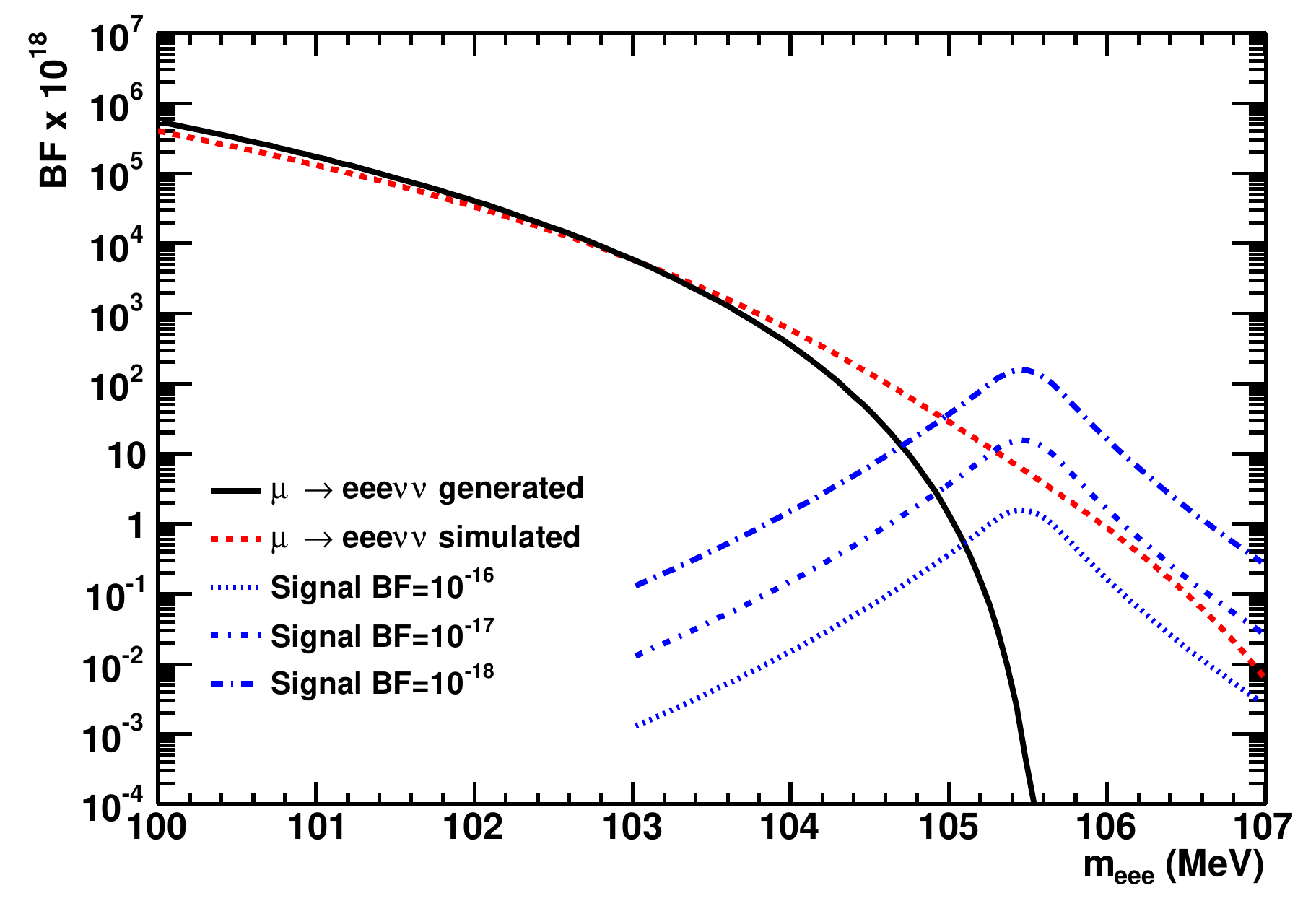}
\end{center}
\caption{The $\mu^+ \rightarrow e^+e^-e^+\nu_e \bar\nu_\mu$ branching fraction before and after convolution with the detector resolution overlaid with signal at different branching 
fractions. Results are shown for 50$\mu$m thick silicon sensors (left) and 30$\mu$m thick silicon sensors (right).}
\label{Fig::mu3e3}
\end{figure}

To estimate the background contributions under these running conditions, we define a signal window as $ 104.9 < m_{eee} < 106.5 ~\Mev$, containing approximately 90\% of the signal. The irreducible background arises from $\mu \rightarrow e^+e^-e^+ \nu\bar\nu$ events where the two neutrinos carry almost no energy. We estimate its contribution to be about 8 events by convolving the branching fraction with the resolution function and integrating in the signal region, as shown in Figure~\ref{Fig::mu3e3}. However, this background depends strongly on the tail of the mass distribution, and small improvements translate into large background reductions. For example, decreasing the thickness of the silicon sensors and the supporting kapton structure by 20\% (40\%) reduces the background down to $\sim 4$ ($\sim 1$) events. Additional improvements of the reconstruction algorithms might further improve the resolution and reduce this contamination as well.

We consider accidental backgrounds produced by the combination of a Michel decay and a radiative Michel decay (2M$\gamma$ decays), or three simultaneous Michel decays (3M decays), where one of the the positrons is misreconstructed or produces an electron by interacting with the detector. In both cases, we assume the decays occurs within the same pixel in the active target, and during the same time window. This yields position and time suppression factors $\delta S = 7.8\times 10^{-7}$ and $\delta t = 2.5\times 10^{-10}$, respectively. The number of background events per second can be expressed as:
$$N_{2M\gamma} = {R_\mu}^2 \delta S \delta t {B(\mu^+ \rightarrow e^+ \nu_e \bar\nu_\mu)}^2 B(\mu^+ \rightarrow e^+ \nu_e \bar\nu_\mu \gamma) P(\gamma \rightarrow e^+ e^-)  P_\mu  \simeq 0.33 P_\mu$$
$$N_{3M} = {R_\mu}^3(\delta S)^2 {B(\mu^+ \rightarrow e^+ \nu_e \bar\nu_\mu)}^3 (\delta t)^2 P_\mu \simeq 0.02 P_\mu$$
where $P(\gamma \rightarrow e^+ e^-)\sim 0.18\%$ is the probability of photon conversion in the target and $P_\mu$ denotes the probability to reconstruct a muon candidate after all selection criteria are applied. The factors $P_\mu$ are estimated by Monte Carlo simulation using the matrix element and differential decay width given in Ref.~\cite{Kuno:1999jp,Djilkibaev:2008jy}. Values of $P_\mu$ of the order of ${\cal O}(10^{-8})$ (${\cal O}(10^{-9})$) are found for 2M$\gamma$ (3M) decays. Both backgrounds are estimated to be less than one event. A similar background level is expected from combinations of $\mu^+ \rightarrow e^+ \overline{\nu}_{\mu} \nu_e$ and $\mu^+ \rightarrow e^+e^-e^+ \overline{\nu}_{\mu} \nu_e$ decays. 

In summary, we have outlined the requirements needed to improve the projected sensitivity of the Mu3e experiment by an order of magnitude using a compact silicon tracker surrounding an active target. We estimate that a stopped muon rate of ${\cal O}(8\times 10^{9})$ would be required to achieve a SES of $5\times 10^{-18}$ for a 3-year run with 100\% DAQ efficiency. Relatively modest improvements on the resolution are needed to maintain the irreducible background at an appropriate level, while an active target proves to be essentially in the reduction of accidental backgrounds.

%% file: conclusion.tex
\section{Conclusions}
\label{conclusions}

With plausible improvements in photon energy resolution provided by measuring the photon in the decay $\mu \to e \gamma$, time resolution and vertex location, it appears feasible to substantially improve the sensitivity of searches for this decay, provided that a sufficiently intense surface muon beam, such as that being studied in the context of Project~X can be provided. The use of an active target and silicon tracking can similarly improve the sensitivity of searches for the rare decay $\mu \to 3e$.

Improvements in the sensitivity of searches for both $\mu \to e \gamma$ and $\mu \to 3e$ decays beyond those in proposed in the MEG upgrade and Mu3e experiments are are well-motivated and appear to be quite possible. To achieve this improvement, will be necessary to improve the experimental resolution in the directions explored herein, and to develop a more intense surface muon beam.

\section{Acknowledgments}
\label{ack}

We acknowledge helpful discussions with Fritz Dejongh and Franco Grancagnolo.
This work was supported in part by the US Deparment of Energy under grant DE-FG02-92ER40701.